\documentclass{article}

\PassOptionsToPackage{authoryear,comma}{natbib}
\usepackage{neurips_2024}

\usepackage{graphicx}
\usepackage[utf8]{inputenc} 
\usepackage[T1]{fontenc}    
\usepackage{hyperref}       
\usepackage{url}            
\usepackage{booktabs}       
\usepackage{amsfonts}       
\usepackage{nicefrac}       
\usepackage{microtype}      
\usepackage{xcolor}         
\usepackage{dsfont}
\usepackage{amsmath}
\usepackage{afterpage}
\usepackage[affil-it]{authblk}
\usepackage{textcomp}
\usepackage{float}

\title{From Transformer to Biology: A Hierarchical Model for Attention in Complex Problem-Solving}

\author[1]{\textbf{Zhongqiao Lin}\textsuperscript}
\author[1,2]{\textbf{Yunwei Li}\textsuperscript}
\author[1]{\textbf{Tianming Yang}\textsuperscript{\dag}}

\affil[1]{Institute of Neuroscience, Key Laboratory of  \authorcr
	 Brain Cognition and Brain-inspired Intelligence Technology, \authorcr
	Center for Excellence in Brain Science and Intelligence Technology, \authorcr
	Chinese Academy of Sciences, Shanghai, China}
\affil[2]{University of Chinese Academy of Sciences\authorcr
	\texttt{\{zqlin, liyw2021, tyang\}@ion.ac.cn}}

\renewcommand*{\thefootnote}{\arabic{footnote}}

\begin{document}

\maketitle
\renewcommand\thefootnote{\fnsymbol{footnote}}
\footnotetext[1]{Corresponding author.}
\nolinenumbers 
\begin{abstract}
Attention is fundamental to cognition, yet it remains a challenge to understand attention in tasks approaching real-world complexity. Here, we approached this problem by modeling gaze patterns of monkeys playing Pac-Man. We first show a transformer network trained to reproduce their gameplay developed internal attention patterns closely matching the monkeys' eye movements. By dissecting the network's attention, we revealed a hierarchical structure comprising two components: a value-based layer encoding fixed object salience, coupled with a dynamic interaction layer tracking relational information between game elements. We further developed a condensed model in which reward-driven attention serves as a gain modulator and is integrated with spatial attention maps, predicting attention as well as the transformer. Together, our study pioneers the use of AI architectures as analytical tools and bridges mechanistic interpretability with cognitive neuroscience to yield novel, testable insights into how the brain coordinates reward, spatial cognition, and attention in complex environments.
\end{abstract}

\section{Introduction}

Attention is a critical cognitive function, as we navigate environments saturated with information that far exceeds the brain's processing capacity \citep{posner1980orienting, posner1980attention}. It serves as a filter, dynamically selecting and prioritizing information to guide adaptive behavior.  Classical frameworks distinguish between bottom-up attention, driven by sensory salience via parietal and temporal cortices, and top-down attention, modulated by prefrontal and frontal eye field control (for review, see e.g., \citet{knudsen2007fundamental, katsuki2014bottom, maunsell2015neuronal, martinez2022visual}). However, this dichotomy has been challenged. For example, the complex role of reward in attention allocation can both align with goal-directed behavior, suggesting top-down control, and operate independently of current task demands, resembling bottom-up processes \citep{awh2012top, theeuwes2024attentional,wolfe2017five}. Especially in a complex, dynamic, and naturalistic setting, the classic framework is too limited for understanding how the brain combines external sensory inputs and task context to direct its attention \citep{draheim2022role}. 

Traditional normative modeling works are hypothesis-driven and rely on heuristics or insights from simplified experimental paradigms \citep{itti2000saliency, torralba2006contextual, judd2009learning, callaway2021fixation}. While interpretable, these models are unable to scale up to model attention in more naturalistic and complex problem-solving scenarios where multiple cognitive demands continuously reshape attentional priorities. Conversely, data-driven approaches using deep neural networks have achieved remarkable success in predicting gaze patterns even in complex scenarios \citep{assens2017saltinet, huang2018predicting, sun2019visual, wang2024transgop}, but they typically function as black boxes, offering little mechanistic insights. Recent hybrid approaches that inject heuristic biases into neural network architectures (e.g., \citet{hahn2018modeling, li2023modeling}) are often constrained by their a priori assumptions, potentially missing critical computational principles underlying biological attention.

\begin{figure}[htb]
    \hspace{-0.45in}
        \includegraphics[width=1.2\textwidth]{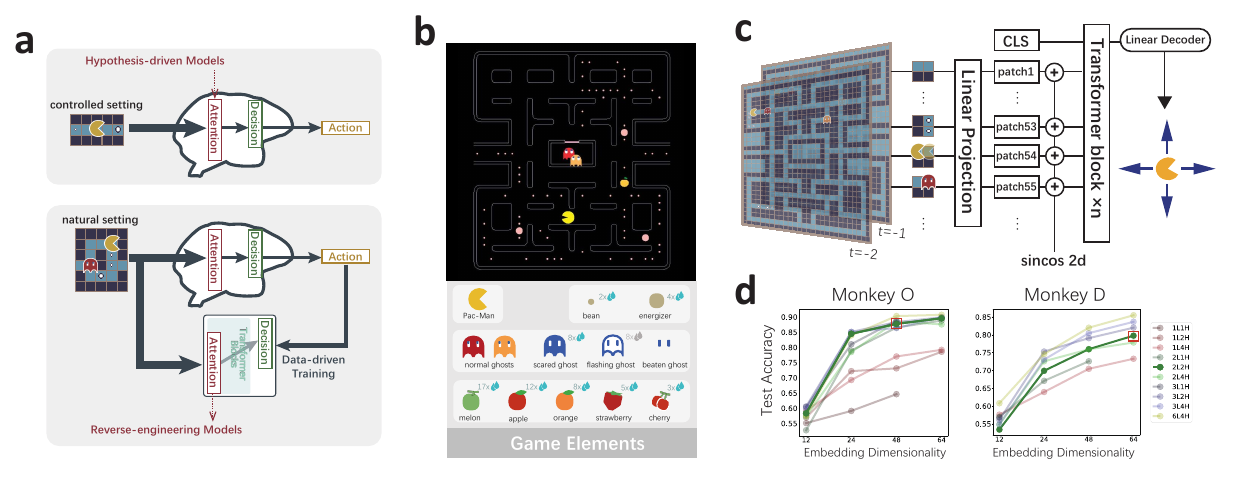}
    \caption{\textbf{Study design and vision transformer model.} \textbf{a.} Two approaches of building attention model. Top: Hypothesis-driven approach. Normative modes based on a small set of variables are based on certain hypotheses and tested in well-controlled experiments. Bottom: Reverse-engineering approach. Neural network models are trained to align its behavior with experimental data, and their internals exhibit functional characteristics of the brain. The models are then reverse-engineered to provide mechanistic interpretation underlying the attention. \textbf{b.} Adapted Pac-Man game. Monkeys received juice rewards, the size of which was indicated in the right-top corner for each object shown below. Two ghost were in the game, and they incurred a time-out penalty. \textbf{c.} Transformer network model. The network is trained to predict the monkeys' directional choice at a junction, and its inputs are two game frames before Pac-Man reaches the junction. \textbf{d.} Determining model hyperparameters. We tested model performance with different numbers of layers,  numbers of heads, and dimensionality of embedding tokens. Balancing the performance and network simplicity, we chose the model with 2 layers, 2 attention heads, 48 token dimensions (monkey O) and 64 token dimensions (monkey D), indicated by the red frame. }
\label{fig:schem}
\end{figure}

Here, we propose a different approach that uses AI models not merely as predictive tools, but as substrates for discovering normative principles of cognition. This approach leverages recent breakthroughs in mechanistic interpretability—techniques for reverse-engineering the computational mechanisms within AI systems \citep{xie2021explanation, min2022rethinking, bereska2024mechanistic, rai2024practical}—and applies them to decode biological cognitive processes. This novel synthesis allows us to extract interpretable, normative models from AI systems that have learned to solve complex tasks, effectively letting the data reveal the underlying computational principles rather than imposing them a priori (Figure \ref{fig:schem}A).

Adopting this approach, we investigated the gaze patterns in monkeys that were trained to play Pac-Man,  a game demanding continuous attention allocation amid competing objectives, mirroring real-world primate behaviors like foraging and predator avoidance mediated by visual and attentional brain networks. We trained transformer networks \citep{vaswani2017attention, dosovitskiy2020image} to play the game as the animals and found that the transformers' attention mechanisms spontaneously aligned with the monkeys' gaze patterns. 

By systematically dissecting the transformer network's attention mechanism, we uncovered a hierarchical computational structure that governs attention allocation, akin to layered neural processing in attention circuits involving bottom-up saliency and top-down modulation \citep{katsuki2014bottom}. Based on these results, we further constructed a normative model that is fully interpretable and maintains high predictive accuracy, achieving what neither traditional normative models nor black-box AI could accomplish alone. Our results suggest that the attention emerged from a static value-driven salience signal integrated with an task-dependent spatial map of objects, with multiplicative gain modulations for coordinating different reference frames during real-time navigation and problem-solving.

\section{Results}
\label{R}
\subsection{Attention rollout}
\label{R1}

Our neural network models were based on standard vision transformers (ViT) \citep{dosovitskiy2020image} that were trained to learn the monkeys' joystick decisions when Pac-Man was at a junction. The monkey data were from a previous study in a published paper \citep{yang2022monkey}. The dataset included the eye movements and joystick movements of two well-trained rhesus monkeys playing an adapted Pac-Man game. Only the joystick movements were used in training. Balancing between the network's performance and simplicity, we chose the model with 2 layers, 2 attention heads, and 48 token dimensions (monkey O) and 64 token dimensions (monkey D) for the rest of the study (See \hyperref[M]{\bf Methods}). The trained network reached a performance of 87.664\% and 79.711\% predicting the two monkeys' choices at junctions, respectively. 
 
\begin{figure}[htbp]
     \hspace{-0in}
        \includegraphics[width=1.0\textwidth]{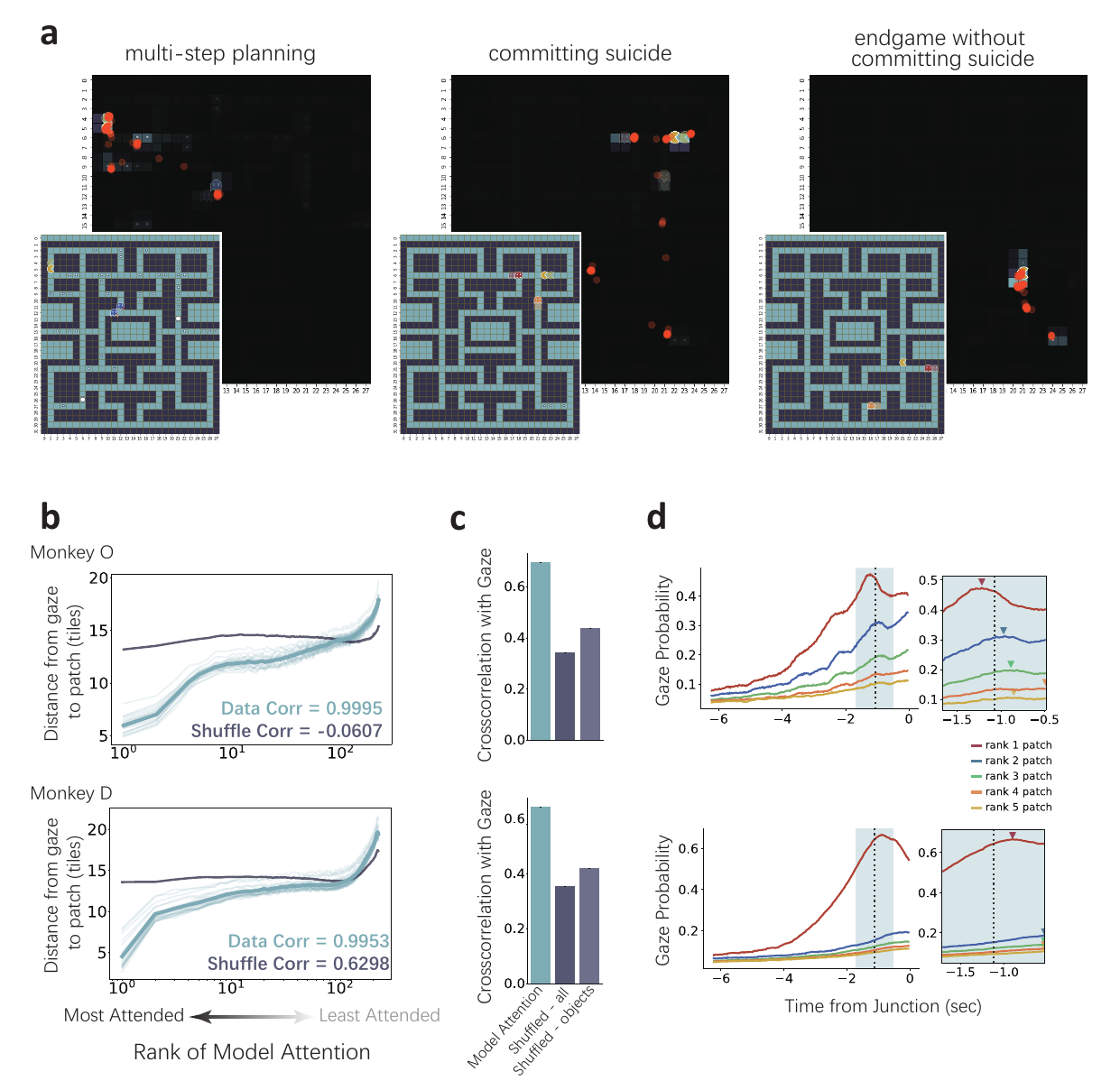}
    \caption{\textbf{The network's attention scores predict the monkeys' gaze patterns.} \textbf{a.} Example games. The network's attention rollout scores (indicated by the brightness at each maze location) closely match the monkeys' gaze locations (red dots), even during complex planning sequences. Left: as Pac-Man approached a junction with two branches ahead, the monkey's gaze was fixated on a scared ghost further away, rather than on the immediate branches. This gaze behavior may indicate the monkey's plan to hunt the ghost. Middle: when Pac-Man was far from the remaining dots and the respawn spot, the monkey committed \textit{suicide} to take advantage of the respawn. We observed that the monkey's gaze alternated between the ghost and the pellets near the respawn spot, suggesting that it was planning its \textit{suicide} strategy. Right: when the remaining dots were near Pac-Man, the monkey ignored the ghosts. See Figure \ref{fig:more example} in the \hyperref[app:appendix]{\bf Appendix} for more scenarios. \textbf{b.} The patch's attention score rank order is correlated with the distance of the monkeys' gaze location from the patch (top: monkey O; bottom: monkey D). Each thin gray line is a testing session, and the light blue is the session average. The error bars (standard error of the mean) are too small to be visible. p-value is under the machine precision. The dark blue is the shuffle among samples. \textbf{c.} The peak value of cross-correlation between the attention score and the gaze pattern is significantly greater than the correlation computed with the shuffled data (top: monkey O; bottom: monkey D). The middle bar is the data shuffled across all maze patches, and the right bar is the data shuffled only among patches with game objects. The error bars (standard error of the mean) are too small to be visible. \textbf{d.} Gaze probabilities to the top-5 patches based on their attention rollout scores (top: monkey O; bottom: monkey D). Patches with higher attention rollout scores not only are associated with higher gaze probabilities but also peak earlier. Dashed line is the averaged joystick movement time and the shaded area is the 1.2 sec time window around the joystick movement.}
\label{fig:stats}
\end{figure}

We examined the networks' attention with the attention rollout metrics \citep{abnar2020quantifying} and observed interesting patterns. Figure \ref{fig:stats} shows three examples. In these examples, we plotted the monkeys’ gaze locations during a 3-tile time-span before Pac-Man entered a junction, along with the network's attention on the same maze. The chosen time-span was when the monkeys were looking around the maze to make a decision on which direction to turn at the junction. In the first example (Figure \ref{fig:stats}A, left), Pac-Man was facing two branches at the junction. In addition to weighing the available rewards in both branches, the monkey was also looking at a scared ghost far away from Pac-Man, suggesting that it was planning to hunt it. The network attention captures this long-range planning. The second example (Figure \ref{fig:stats}A, middle) involves an interesting strategy that the monkey learned to use, namely, \textit{suicide} \citep{yang2022monkey}. Here, the monkey would choose to run into a ghost and die when Pac-Man was far away from the respawn spot, and the remaining pellets were near the respawn spot. Doing so would allow the monkey to collect the remaining pellets more easily after the respawn. In this example, the monkey was looking at both the ghost and the pellets around the respawn spot, reflecting that it was planning a \textit{suicide}. Again, the network's attention captures the same pattern. These two examples suggested the ghosts were important for attention. Yet, in the final example, the monkey ignored them. Here, Pac-Man was close to the only remaining pellets in the maze, and both the monkey and the network ignored the ghosts and focused their attention on the remaining pellets, revealing that the attention to the ghosts were modulated by the game context (Figure \ref{fig:stats}A, right). More examples are included in Supplementary Figure \ref{fig:more example}.

Overall, the monkeys' gaze patterns reflected their complex game play, which are closely matched by the attention in the transformer network. To quantify the similarity between the networks' attention and the monkeys' gaze, we divided the maze into patches and ranked them by their attention rollout scores. Patches receiving higher attention rollout scores were located closer to the monkeys' actual gaze locations (Figure \ref{fig:stats}B, Spearman correlation, corr=0.9995 (monkey O) and 0.9953 (monkey D), \(p\) is under the machine precision). The same conclusion holds when only the non-empty patches are used in the analyses (Figure \ref{fig:more example}).

We further validated this relationship through three complementary analyses. First, we found that the probability of the gaze falling on a patch (see \hyperref[M]{\bf Methods}) increased by 23.14\% (monkey O) and 45.14\% (monkey D) from the lowest attention rank to the highest attention rank, while the shuffled data only increased by 1.27\% and 0.68\% in the two monkeys, respectively.  Second, an 2D cross-correlation analysis revealed strong spatial correspondence between attention rollout score maps and gaze heatmaps. The observed correlations significantly exceeded those from control conditions where attention scores were shuffled either across all patches or only among object-containing patches (Figure \ref{fig:stats}C). Finally, we observed a temporal alignment: patches with higher attention ranks were examined first, reaching their peak gaze probability earlier (Figure \ref{fig:stats}D). This is consistent with the intuition that more important information should be collected first. Together, these results demonstrate that transformer attention can well captures both the spatial and the temporal pattern of the monkeys' gaze allocation.

Moreover, to determine whether the transformer learned monkey-specific attention patterns or merely reflected general game mechanics, we compared its attention to human judgments. Ten human participants viewed 100 game states from the transformer's training data and marked tiles they considered most relevant for decision-making. The participants had no knowledge of the monkeys' game play, therefore their choices mostly reflected the game mechanics. The human judgments showed high inter-subject agreement but diverged significantly from the transformer's attention patterns (Jaccard similarity analysis, Supplementary Figure \ref{fig:hum exp}). Importantly, when predicting actual monkey gaze locations, the transformer outperformed the humans, while a random baseline performed substantially worse than both. The results confirm that the transformer trained to mimic the monkeys' gameplay captured the attention patterns specific to individual monkeys. The model learned not just where one should look based on game mechanics, but where these particular monkeys actually directed their attention during their decision-making.

\subsection{Interpreting the model's attention}
\label{R2}
Next, we aimed to reveal the computation in the attention mechanism in our model.

We first expanded the equation for attention rollout score by layers. As the network's final output is only based on the CLS token, we only examined the CLS token's attention, which is the first row of the attention matrix. We omit the head notations for simplicity:

\begin{equation}
    \tilde{a}_{CLS}=0.25({\bf A}^{(1)}_{1,2:}+{\bf A}^{(2)}_{1,2:}+{\bf A}^{(2)}_{1,:}{\bf A}^{(1)}_{:,2:})\text{,}
    \label{eq layer attn}
\end{equation}

where \({\bf A}^{(1)}\) and \({\bf A}^{(2)} \in \mathbb{R}^{225\times225}\) are the attention score matrices of the first and the second layer.

\begin{figure}[htbp]
    \hspace{-0.5in}
        \includegraphics[width=1.15\textwidth]{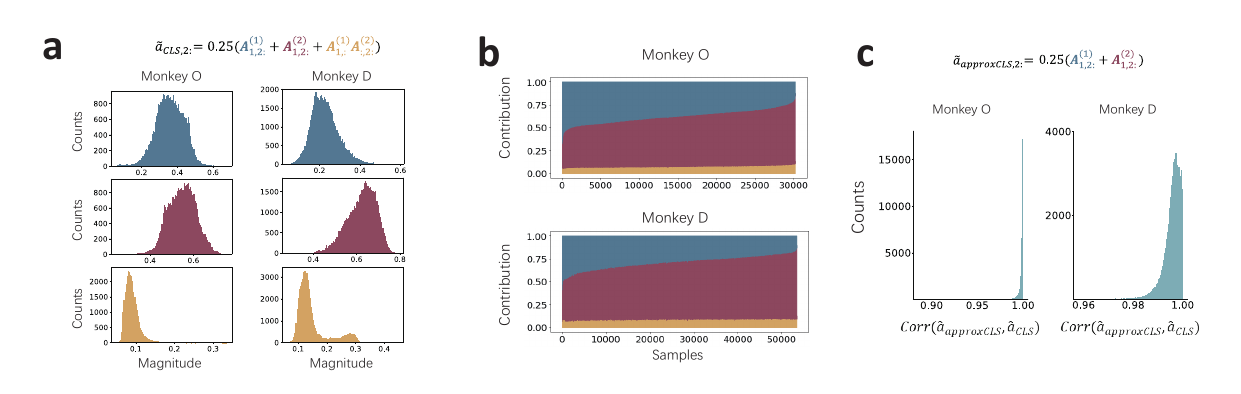}
    \caption{\textbf{Attention rollout score can be approximated by only 2 components.} \textbf{a.} The rollout score of the CLS token to the maze patch tokens is the sum of 3 components: the first row of layer 1 attention, the first row of layer 2 attention, and the multiplication between layer 1 attention and layer 2 attention. The distributions of the vector norm of these components are plotted for both monkeys (left: monkey O; right: monkey D), suggesting the multiplication component is small \textbf{b.} The ratio of the vector norm of the 3 components. The samples are sorted by the ratio of the layer 2 attention. \textbf{c.} The rollout score can be well approximated by only the first two components. The distribution of Pearson correlation between the approximation and the original rollout score is plotted. Only patches with game objects are involved in the computation of correlation.}
\label{fig:expand}
\end{figure}

We noticed that the interaction term can be ignored. First, the norm of the multiplication between \({\bf A}^{(1)}\) and \({\bf A}^{(2)}\) was much smaller than those of the other two attention vectors (Figure \ref{fig:expand}A). In addition, the interaction term contributed only a small proportion in all samples (Figure \ref{fig:expand}B), and removing it produced nearly identical attention rollout scores (Figure \ref{fig:expand}C). Therefore, we focused on the remaining two attention vectors.

\subsubsection{First layer}

The attention score of the first layer is:

\begin{equation}
    {\bf A}^{(1)}_{1,2:} =softmax(\frac{\alpha^1_{CLS,2:}}{\sqrt{d_k}})\text{,}
\end{equation}

\begin{equation}
    \alpha^1_{CLS,2:} = LN(x_{CLS}){\bf W}^1_Q{\bf W}_K^{1T} LN({\bf X}_{2:,:}^T){,}
\end{equation}

where \(x_{CLS}\) is the learnable CLS token, \({\bf W}^1_Q\), \({\bf W}^1_K \in \mathbb{R}^{d\times{d/h\cdot h}}\) are the query and the key weight matrix of the first layer, respectively. \(LN(x)\) is the layer norm computation in the first layer. The layer norm computation is nonlinear and makes the interpretation difficult, but it can be removed with extra finetuning on the trained model \citep{heimersheim2024you}. The finetuned model maintains both high choice accuracy and a strong correlation with the original layer 1 attention patterns (Supplementary Figure \ref{fig:finetuning}). After the nonlinear layer norm removed, the layer 1 attention becomes:

\begin{equation}
    \alpha^{1,*}_{CLS,2:} = x^*_{CLS}{\bf W}^{1,*}_Q{\bf W}_K^{1,*T} {\bf X}_{2:,:}^{*T}
    = x^*_{CLS}{\bf W}^{1,*}_Q{\bf W}_K^{1,*T} {\bf W}_{emb}^{*T} {\bf S}_r^T+x^*_{CLS}{\bf W}^{1,*}_Q{\bf W}_K^{1,*T} {{\bf E}}_{pos}^{T}\text{,}
\end{equation},

 where * represents the parameters and variables from the linearized model, \({\bf W}^*_{emb} \in \mathbb{R}^{136\times{d} }\) is the linear embedding matrix, \({\bf S}_r \in \mathbb{R}^{224\times{136}}\) is the reshaped input game state, and \({\bf E}_{pos}\) is the positional embedding matrix. As the positional encoding and the maze channels in the input encode maze information but are otherwise invariant in the game, \(\alpha^{1,*}_{CLS,2:}\) can be further rewritten as the sum of a maze component and an object-dependent component:

\begin{align}
    \alpha^{1,*}_{CLS,2:} &=x_{CLS}{\bf W}^{1,*}_Q{\bf W}_K^{1,*T} {\bf W}_{maze}^{*T}{\bf S}_{r, maze}^T+x^*_{CLS}{\bf W}^{1,*}_Q{\bf W}_K^{1,*T} {{\bf E}}_{pos}^T+x^*_{CLS}{\bf W}^{1,*}_Q{\bf W}_K^{1,*T} {\bf W}_{obj}^{*T} {\bf S}_{r, obj}^T \nonumber\\
    \nonumber \\
    &={\bf \alpha}^*_{maze}+x^*_{CLS}{\bf W}^{1,*}_Q{\bf W}_K^{1,*T} {\bf W}_{obj}^{1,*T}{\bf S}_{r,obj}^T\text{,}
\end{align}

where \({\bf W}^*_{maze}\in\mathbb{R}^{8\times{d}}\),\({\bf W}_{obj}\in\mathbb{R}^{128\times{d}}\) are sub-matrices that consist of the rows corresponding to the maze channels and the other object channels of the embedding matrix \({\bf W}^*_{emb}\), respectively. \( {\bf S}_{r, maze}\) and \( {\bf S}_{r, obj}\) are sub-matrices of the corresponding dimensions of \( {\bf S}_r\). \({\bf \alpha}^*_{maze}\) is the maze component of the CLS token attention and reflects the attention bias within the maze.

\begin{figure}[htbp]
    \hspace{-0.4in}
        \includegraphics[width=1.15\textwidth]{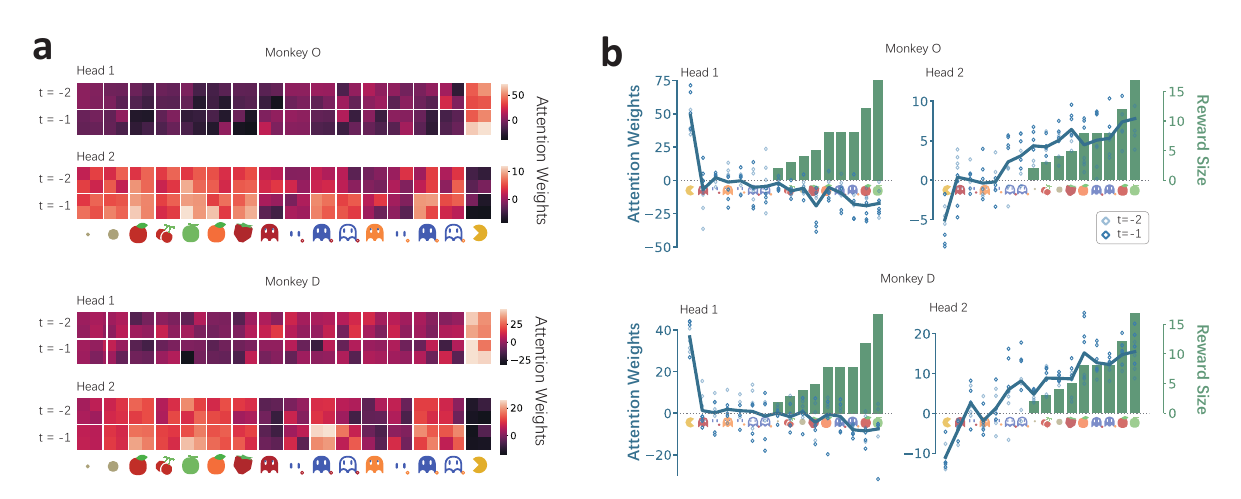}
    \caption{\textbf{Layer 1 attention is driven by salience.} \textbf{a.} The attentional weights of the linearized layer 1 transformer. The weights are arranged to match the four tiles in a patch for each game object and each time frame. \textbf{b.} The attentional weights of the layer 1 transformer, plotted against the reward magnitude of the game objects (left: head 1, right: head 2). The curves are the average weights of each object, and the bars are the reward size (value) of the objects. Diamonds are the object weights plotted separately for each tile and each frame (first frame: light blue; second frame: dark blue), as shown in \textbf{a}.}
\label{fig:l1attn}
\end{figure}

We examined separately the two heads' attention weights on the game objects based on the second term \(x^*_{CLS}{\bf W}^{1,*}_Q{\bf W}_K^{1,*T} {\bf W}_{obj}^{1,*T}\) (Figure \ref{fig:l1attn}A), which is a \(1\times{136}\) vector. In head 1, the attention weights of Pac-Man are predominantly high, and the contributions from the other game objects are negligible. Therefore, the attention in head 1 is mostly directed toward Pac-Man. In contrast, the head 2's attention weights are high for objects that yield rewards. They include pellets, energizers, fruits, and scared ghosts, but not ghosts in the normal or the dead mode. The attention correlated with the actual reward associated with the objects in head 2 (Figure \ref{fig:l1attn}B). This value-based attention is similar to what has been demonstrated in the brain \citep{zhang2022reward, anderson2013value}. Together, the layer 1 attention of the original model depends on the location of the Pac-Man and the subjective values of the game objects.

The maze component \({\bf \alpha}^*_{maze}\) in the two heads contributes little to the overall attention  (Supplementary Figure \ref{fig:finetuning}). We ignore this component in the further reconstructions.

Overall, the first-layer attention is driven by where Pac-Man is (head 1) and what the non-Pac-Man objects' reward values are (head 2). This is similar to the type of attention termed bottom-up in cognitive science to describe the attention driven by sensory and reward salience.

\begin{figure}[htbp]
    \vspace{-0.8in}
        \includegraphics[width=1.\textwidth]{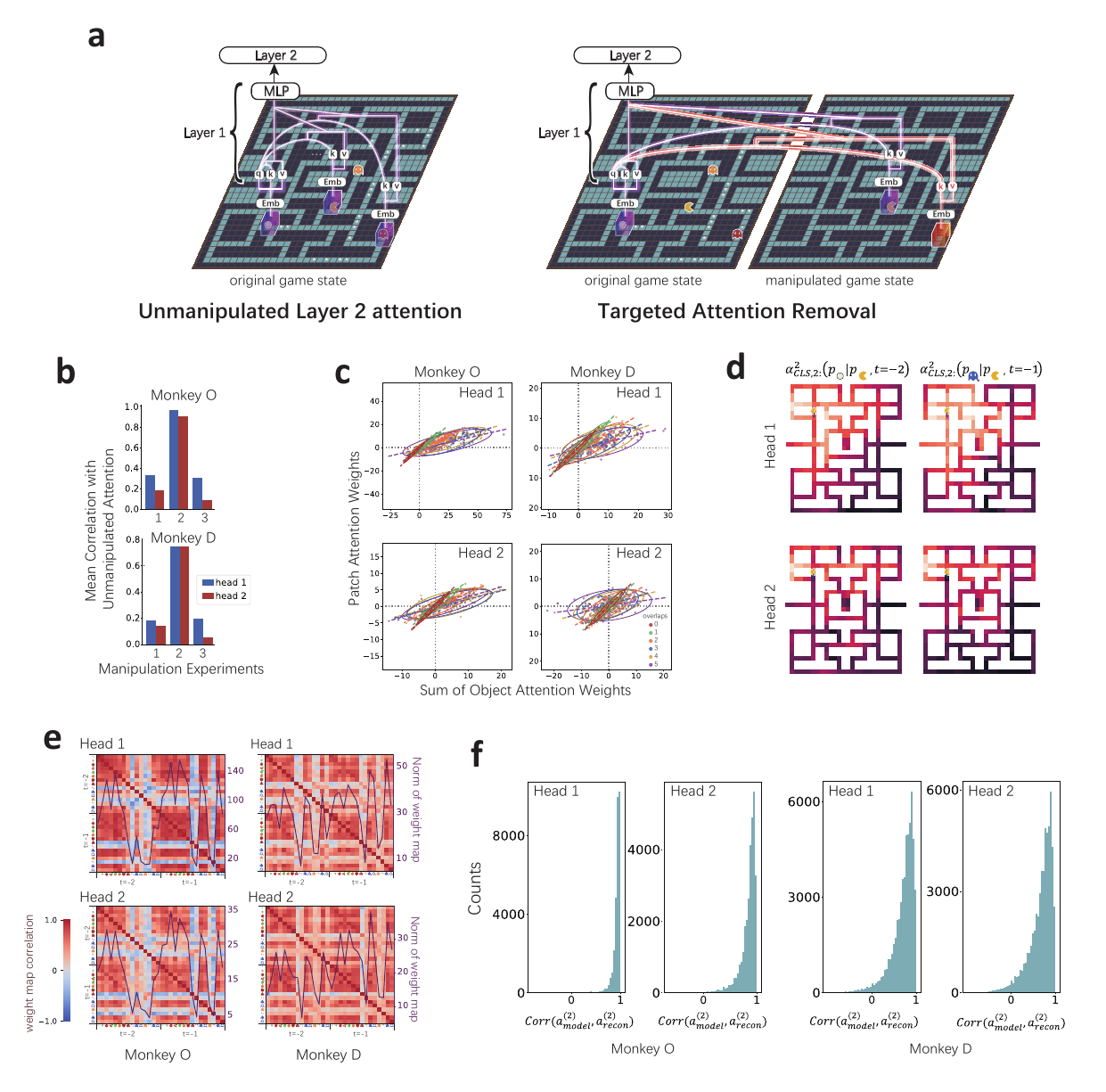}
    \caption{\textbf{Layer 2 attention is a Pac-Man-centric attentional map. a.} Target attention removal. The schematics show the computation of the layer 2 attention from the classification token to an energizer patch. Normally (left), the energizer patch sends its query vector to all board patches with the matching key vector.  In target attention removal (right), a patch's query interacts with its own key and value vectors, as well as those of all other patches computed from a manipulated input where particular target objects have been removed. Thereby, the information from the target objects is blocked selectively from the attention computation of the current patch, while the information from the other objects is intact. \textbf{b.} Pac-Man-object interaction is important for attention. Plotted are the correlations between the manipulated layer 2 attention and the original layer 2 attention in three 3 experiments. In experiment 1, we removed all the objects from the manipulated game state. In experiment 2, we kept only the Pac-Man. And in experiment 3, we shuffled the position of the Pac-Man and removed all other objects. \textbf{c.} Overlapping objects' attention contribution was discounted. Each data point is a patch, and they are grouped by the number of overlapping objects, indicated with color. 0 means no overlapping objects on the tile. The horizontal axis indicates the sum of objects' attention weights computed from the non-overlapping condition. The vertical axis indicates the actual patch attention weights. For clarity, only 100 data points are shown for each group. The ellipses are the contours of estimated Gaussian distributions of each group, and the dashed lines are their major axes. Smaller angles indicate more discounting. \textbf{d.} Example Pac-Man-centric weight maps of the game objects. The locations of the Pac-Man are marked on the maps. \textbf{e.} Similarity between Pac-Man-centric maps of different objects. Color indicates correlation between different objects' maps. Purple curve indicates the contribution to the total attention of the objects, measured by the norm of the map. \textbf{f.} The distributions of the accuracy from the shared map model, measured by the correlation between the original layer 2 attention and the reconstructed layer 2 attention.}
\label{fig:l2attn}
\end{figure}


\subsubsection{Pac-Man object interaction}
An important part of the Pac-Man gameplay is the relationship between Pac-Man and the other game objects, such as the relative locations of the pellets, the ghosts, etc., which should be captured by the network's attention. This does not happen in layer 1, where the attention associated with the CLS token is only computed between CLS and the game patches, and thus does not capture the interactions between game objects in different patches. In contrast, in layer 2, each token contains the information from the other tokens through the attention mechanism of layer 1 (Figure \ref{fig:l2attn}A, left), providing the attention of the CLS token in layer 2 the opportunity to capture the interaction between game objects. 

To understand how the interactions between the game objects contribute to the layer 2 attention, we introduce an ablation procedure called targeted attention removal (Figure \ref{fig:l2attn}A, right). This technique selectively disrupts specific object interactions by manipulating the attention computations in layer 1. The procedure works as follows. Given a game state, we create two versions—the original and one with specific objects removed. We compute the embeddings for both versions separately. Then, we calculate a variant of cross-attention between the original version and the manipulated companion, so that each patch no longer receives information from removed objects, while information flow from the patch itself and the remaining objects are intact. By propagating these modified layer 1 attention through to layer 2, we can investigate how the absence of specific object interactions, but not the absence of objects themselves, affects the network's higher-level attention,  allowing us to identify which object relationships are critical for the transformer's decision-making.

We focused on two key interaction types: the interactions between Pac-Man and other game objects, and the interactions among non-Pac-Man objects. To isolate their contributions, we applied targeted attention removal using three manipulated versions of each game state \textbf{S}:

• \({\bf S}_0\): Removed all game objects (including Pac-Man), retaining only the maze structure

 • \({\bf S}_{P}\): Removed all objects except Pac-Man

 • \({\bf S}_{PR}\): Similar to \({\bf S}_{P}\), except Pac-Man relocated to a random junction

We then measured how each manipulation affected layer 2 attention by computing correlations between the original and manipulated attention patterns. High correlations indicate the removed interactions were dispensable for the network's computation, while low correlations reveal interactions critical for decision-making.

The results reveal a striking pattern: layer 2 attention primarily encodes how other game objects relate to Pac-Man's position (Figure \ref{fig:l2attn}B). When we disrupted Pac-Man-related interactions (\({\bf S}_0\) and \({\bf S}_{PR}\)), attention correlations dropped dramatically. However, removing interactions between non-Pac-Man objects (\({\bf S}_{P}\)) left correlations virtually unchanged. Therefore, layer 2 implements a Pac-Man-centric attention mechanism—each object's attention weight depends solely on its relationship to Pac-Man, not on relationships with other objects. Crucially, this attention is context-dependent: relocating Pac-Man (\({\bf S}_{PR}\)) disrupts the pattern, confirming that the network computes genuine spatial relationships rather than static object features.

\subsubsection{Overlapping objects' attention contribution is discounted}
Next, we investigated how overlapping objects on the same tile contribute to that patch's raw attention score. We first computed each object's contribution to attention when there were no other objects occupying the same tile, which is the difference between a patch with the object alone and when the patch is empty. When there were multiple objects on the same tile, we found that the patch's total attention was not just the linear sum of its constituent object attentions, but the sum scaled by a discounting factor (Figure \ref{fig:l2attn}C). When two objects are together, each object contributes less than when it is alone, and when there were three objects, their contribution were even less, and so on. As more objects occupied the same tile, each object's contribution was discounted further. 

To capture this relationship, we constructed a group-specific linear model, where scaling coefficient decreases with the number of overlapping objects in a patch. This model accurately reconstructed patch attention weights. Alternative models—such as winner-take-all (using only the highest-weighted object) or soft prioritized occlusion—performed significantly worse (see Supplementary Figure \ref{fig:layer 2 supplementary}).

\subsubsection{A conjunctive spatial and value attention template}
Having decomposed patch attention into individual object contributions, we examined the spatial patterns of attention each object type evoked across the maze. Figure \ref{fig:l2attn}D displays the example attention maps given a particular Pac-Man location for 2 objects, energizer and scared ghost, in the two time frames and the two layer 2 heads, respectively. Indeed, these maps were similar within each attention head in general—when Pac-Man occupied the same position, different objects produced nearly identical spatial attention patterns (Figure \ref{fig:l2attn}E).

This observation suggested that all objects may share a common spatial attention template, modulated with an object-specific gain. To test this hypothesis, we formulated the problem as a bilinear decomposition: attention equals the product of a shared spatial map and object-specific scaling weights. Using an iterative algorithm (see \hyperref[M]{\bf Methods}), we successfully recovered both components. The decomposition accurately reconstructed the original layer 2 attention patterns (Figure \ref{fig:l2attn}F), confirming our hypothesis. We further tried to interpret the shared object maps with several heuristic models (figure \ref{fig:latent map models}). We found that both distance-based models and spatial chunking models explain the map better than chance. 

Importantly, the object-specific gain correlated strongly with each object's reward value (figure \ref{fig:layer 2 supplementary}), revealing an elegant solution: layer 2 combines a spatial template (encoding positions relevant to Pac-Man) with value-based gain modulation (encoding object importance).

These final results complete our mechanistic understanding of the Layer 2 attention. It contains a conjunctive representation of a Pac-Man-dependent, object-based spatial attentional map, modulated by the relative importance of each object, i.e., its value, as a gain factor. This implements a multiplicative code that answers two questions simultaneously: "Where should I look given Pac-Man's position?" (spatial attentional map) and "How important is each object type?" (value-based gain). This efficient representation allows the network to flexibly adapt to different game contexts while maintaining computational simplicity.

\subsection{A condensed model of attention}

Put together, our dissection of the transformer's attention mechanism reveals that it comprises two distinct components. The value-based attention in layer 1 encodes Pac-Man's location and the intrinsic value of each object, and the interaction-based attention in layer 2 captures spatial relationships between Pac-Man and other objects.

Based on this mechanistic understanding, we formulated a condensed model that computes the attention at any tile \(T\) in a given game state \(S\). It is a function of the set of objects on tile \(T\) and the Pac-Man's location \(tile_{pacman}\) (head notions are omitted for simplicity):

\begin{equation} 
    Attention Map = \sigma({\bf L}v_{i}) + \sigma(h(n)({\bf L}u\odot{\bf M})+b)
    \label{eq_condensed_moddel}
\end{equation}.

The model components map directly to our findings :

\begin{itemize}
    \item \textbf{\({\bf L}\)}: A 304 × 32 indicator matrix encoding object presence (32 features = 16 object types × 2 temporal frames across 304 walkable tiles)
    \item \(v\): Layer 1's attention weights, capturing Pac-Man's position (head 1) and object values (head 2)
    \item \(u\): Object-specific scaling factors for the spatial attention template
    \item \({\bf M}\): The shared spatial attention template centered on Pac-Man's location
    \item \textbf{\(h(n)\)}: Competition function that discounts attention by the number of overlapping objects \textit{n}
    \item \(b\): Baseline attention for empty tiles
    \item \(\odot\): Hadamard product
\end{itemize}

\begin{figure}[htbp]
    \hspace{-0.4in}
        \includegraphics[width=1.1\textwidth]{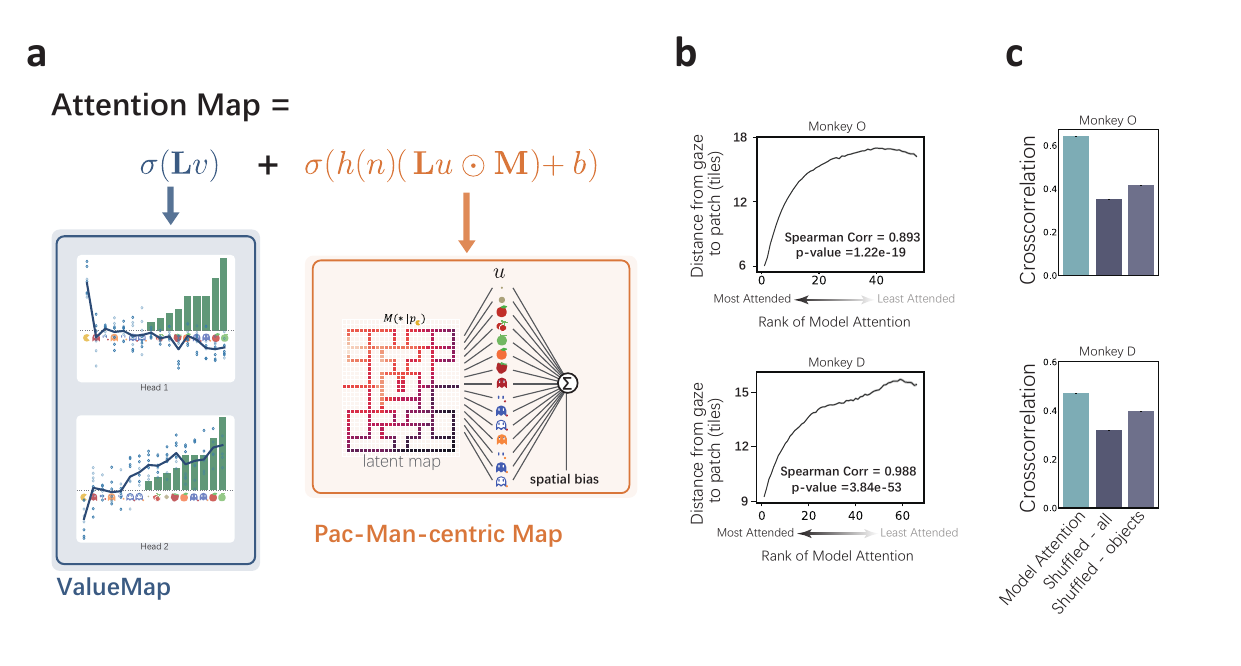}
    \caption{\textbf{The condensed model. a.} The schematic of the two components of the condensed model.  \textbf{b.} The performance of the condensed model. The rank order of the model's predicted attention is correlated with the distance of the monkeys' gaze location from the patch (top: monkey O; bottom: monkey D). Only patches that are non-empty in at least 10\% of the samples are shown. \textbf{c.} The peak value of cross-correlation between the model prediction and the monkeys' gaze position is significantly greater than the correlation computed with the shuffled data (top: monkey O; bottom: monkey D). The middle bar is the data shuffled across all maze patches, and the right bar is the data shuffled only among those with game objects. The error bars (standard error of the mean) are too small to be visible. 
    }
\label{fig:fullmodel}
\end{figure}

Our condensed model successfully captured the monkeys' attention allocation despite its simplified structure. The model's attention scores showed strong correspondence with actual gaze patterns (Figure \ref{fig:fullmodel}B). Cross-correlation analysis further confirmed this relationship. The spatial correlation between the condensed model's attention maps and the monkeys' gaze heatmaps matched that of the full transformer model (Figure \ref{fig:fullmodel}C), demonstrating that our mechanistic decomposition preserved the transformer's predictive accuracy while achieving full interpretability.

In summary, this condensed formulation achieves two goals. First, it provides a fully interpretable model where each attention component has a clear cognitive meaning. Second, while substantially reducing computational complexity relative to the transformer, the condensed model accurately predicts the monkeys' gaze pattern in a naturalistic setting.

\section*{Discussion}
\label{Discuss}
Here, we demonstrated that transformer models trained on behavioral data can capture primate attention allocation in complex, dynamic environments. By systematically dissecting the learned attention mechanisms, we proposed an accurate and interpretable model for the animals' attention during complex problem solving.

Our approach builds upon the power of transformers \citep{vaswani2017attention, dosovitskiy2020image}. We were able to train the transformer with just two layers to capture monkeys' decision making, and they captured monkey-specific attention patterns. Notably, human observers, while consistent among themselves, failed to predict monkey gaze as accurately as the model. The strong correlation between model attention and gaze patterns across multiple temporal and spatial metrics indicates that even small-scale transformers can serve as accurate models of attention dynamics, and their relative simplicity allows for in-depth investigations into their internal mechanisms.

Recent advances in large language models (LLMs) sparked significant interest in exploring parallels between transformer architectures and the brain. For instance, attention scores extracted from LLMs have been shown to align with human eye-tracking patterns recorded when reading the same text \citep{eberle2022transformer, brandl2022every, bensemann2022eye, kewenig2023multimodality}. Additionally, the representations from LLMs modules can be mapped onto brain activities \citep{caucheteux2022brains,schrimpf2021neural, kumar2024shared}. While these studies focus on the similarity between the transformer and the brain, the mechanistic interpretation of the examined representations was often overlooked. Our work bridges this gap and explains how the transformer computes attention.

Our computational decomposition of attention during complex problem-solving map naturally onto distinct neural computations. Layer 1's state-independent encoding of object values reflects the brain's reward circuitry that acquires and represents stimulus-reward associations, likely through interactions between ventral striatum, orbitofrontal cortex (OFC), and posterior parietal regions \citep{schultz2000reward, haber2010reward, womelsdorf2015long, murray2018specializations}. In particular, a recent study reveals that the OFC neurons, similar to layer 1 in our transformer model, maintain persistent value representations of stimuli that remain consistent across different behavioral contexts \citep{zhang2022reward}. This stable value representation captured by the layer 1 component computationally could bias attention and decision-making throughout the brain, providing a bottom-up drive.

Layer 2's dynamic spatial computations align with the spatial priority maps that have been demonstrated in the posterior parietal cortex (PPC), which integrate bottom-up salience with top-down goals, dynamically updating based on current behavioral demands \citep{colby1999space, bisley2010attention, ptak2012frontoparietal}. In addition, layer 2's selective encoding of Pac-Man-to-object relationships may be considered as a sophisticated solution to a fundamental challenge in spatial cognition: coordinating attention allocation in different reference frames \citep{behrmann1999attention, scholl2001objects}. The game environment provides a stable allocentric reference frame—the maze layout and object positions remain fixed in world coordinates. However, the behavioral relevance of each location depends critically on Pac-Man's current position, of which the reference frame is centered around Pac-Man. Layer 2 bridges these reference frames by computing attention weights based on Pac-Man's perspective, which may be considered as egocentric if Pac-Man is considered as a proxy of the agent's ego, and projecting them onto the fixed allocentric spatial map.

This computational strategy mirrors how the brain solves reference frame challenges during navigation and spatial attention. The PPC is involved in the transformations between sensory inputs using different reference frames, such as eye-centric, head-centric, etc, and allocentric spatial representations \citep{andersen1997multimodal, andersen2002intentional, chang2010idiosyncratic}. Our finding that layer 2 exclusively encodes Pac-Man-centered relationships—while ignoring object-to-object relationships—makes computational sense: only the agent's position defines a meaningful reference frame from which to evaluate spatial priorities. The multiplicative gain modulation we observed may implement these coordinate transformations, scaling allocentric location values by their egocentric relevance. 

By factorizing value (layer 1) from spatial accessibility (layer 2), the transformer discovered a solution that parallels how the brain might separate "what is valuable" from "what is reachable from here." This architectural principle could inform our understanding of how biological attention systems efficiently navigate complex environments where multiple reference frames must be coordinated in real-time.

In summary, our work establishes a new paradigm for understanding the neural basis of attention. This approach offers testable predictions about how specific neural circuits contribute to attention allocation during complex tasks, opening avenues for targeted neurophysiological investigations that can validate and refine our computational framework.

\section{Methods}
\label{M}

\subsection{Behavior experiments}
\label{M1}
The monkey behavior experiments and data collection were reported in a previous paper \citep{yang2022monkey}. Briefly, two rhesus monkeys were trained to play an adapted Pac-Man game. They gained juice rewards by clearing all the pellets in a maze and avoiding being caught by ghosts. Taking an energizer turned the ghosts into a scared mode, in which the ghosts could be eaten for additional rewards. During the entire experiment, the eye traces of the monkeys were recorded with an infrared eye-tracking system (EyeLink\textregistered 1000 Plus).

\subsection{Dataset}
\label{M2}
We trained transformer networks to predict the monkeys' directional choice when Pac-Man is at a junction tile. To capture Pac-Man's movement, the input to the networks includes two game frames in which Pac-Man is one and two tiles away from entering the junction. Each frame is a \(32\times28\times17\) tensor, where 32 and 28 are the height and width of the maze in the unit of tiles, and 17 is the number of feature channels ( \hyperref[app:appendix]{\bf Appendix}), each containing a binary indicator for the respective object. The two frames are concatenated into a tensor \({\bf S}\in\mathbb{R}^{32\times28\times34}\).

140 sessions from monkey O and 125 sessions from monkey D were used for training, five held-out sessions from each monkey were used for testing the decision accuracy. The ratio of overlapped samples in the training and the testing set was negligible. For the analyses conducted in this study, 20 and 15 held-out sessions from monkey O and monkey D were used, respectively. These sessions were selected based on the availability of high-quality eye-tracking data, but were independent of the prediction accuracy of the transformer model.

\subsection{Transformer Network}
\label{M3}
The networks are based on standard vision transformers (ViT)\citep{dosovitskiy2020image,beyer2022better}. The input of the model is a \(32\times{28}\times{34}\) tensor, where the last dimension corresponds to 17 feature channels in two frames. The description of each channel can be found in Table \ref{channels}. The input \(S\) is reshaped into a sequence of flattened patches \(s_t\in\mathbb{R}^{2^2\times34}\) where \(2\times2\) is the size of each patch. The flattened patches are mapped to \textit{d} dimensions with a trainable embedding projection. We use fixed 2D sin-cos position embeddings and a learnable classification (CLS) token.

We chose the hyperparameters by varying the number of transformer layers, the number of embedding dimensions, and the number of attention heads. We trained the model on 140 monkey experiment sessions and tested the model on 5 sessions on a held-out test set. Balancing between model simplicity and performance (Figure \ref{fig:schem}C), we chose the model with 2 layers, 2 attention heads, 48 token dimensions (monkey O) and 64 token dimensions (monkey D) for the rest of the study.

We used AdamW as the optimizer \citep{loshchilov2017decoupled}, with an initial learning rate = \(10^{-4}\) and weight-decay = \(10^{-7}\). The batch size is 768. All models were trained on an A100 GPU for 800 epochs. 

To linearize the computation of the first layer attention, we removed the first layer norm operation in the trained transformer model and continued to train the model for 200 extra epochs (\citep{heimersheim2024you}).

\subsection{Attention rollout and statistics}
\label{M4}
Following \citep{abnar2020quantifying}, we computed the attention rollout score as below:
\begin{equation}
\tilde{\bf A}^i=
\begin{cases}
(0.5{\bf A}^i+0.5{\bf I})\tilde{\bf A}^{i-1} &  \text{if}\; i>0\\
(0.5{\bf A}^i+0.5{\bf I}) &  \text{if}\; i=0
\end{cases}
\label{eq rollout}
\end{equation}
where \({\bf A}^i\) is the attention matrix of layer \(i\) summed across all attention heads and \({\bf I}\) is the identity matrix.

To compare the attention of the transformer model and the gaze pattern of the monkeys, we took the first row of the attention rollout score matrix and computed the correlation between the rollout score and the gaze data in two different ways. We computed the Spearman correlation between the distance from gaze positions of each patch and its rollout score. Gaze positions in the time of three tiles before the junction were used for the analysis. The distance between gaze positions and a patch was computed by averaging the distance between each gaze position and its closet tile within the patch. The Spearman correlation and the statistical inference were implemented with scipy.stats module.

To account for the systematic drift of eye data during the recording, we also computed a second type of correlation based on the 2D cross-correlation. We first generated a heat map for the gaze positions to the same resolution as the patches. We then computed the maximum 2D cross-correlation between the attention rollout score and the gaze heatmap with the offset varying from -4 to +4 tiles (Figure \ref{fig:stats}C). The controls were computed similarly, with the attention scores shuffled across all patches or across only patches with objects.

As we only considered the first row of the rollout matrix, i.e. the attention score from the CLS token to the maze patches, we derived equation \ref{eq layer attn} from equation \ref{eq rollout}. To measure the relative contribution of the three terms in equation \ref{eq layer attn}, we computed their vector norms and their ratio to the sum of the three norms in each sample (Figure \ref{fig:expand}). 

Showing that the multiplication term only contributes to all samples tested minimally, we approximated the attention rollout score as:
\begin{equation}
    \tilde{a}_{approxCLS} = {\bf A}^{(1)}_{1,2:}+{\bf A}^{(2)}_{1,2:}
\end{equation}

\subsection{Gaze probability}
In order to calculate how much gaze probability correlates the attention rank, we labeled distances that were less than 3-tiles (\(\approx\)0.9\textdegree) as gazes, and computed the gaze probability by averaging the number of gazes across samples. 

We adopted a consistent criterion when computing the monkeys' gaze probability to the top-5 rollout-scored patches. To identify the peak of each probability curve, we denoised each curve by reconstructing the curve with a set of radial basis functions. The peak is then defined with the denoised curves. 

\subsection{Targeted attention removal}
To narrow down object interactions that contribute to the layer 2 computation (i.e. the interactions in layer 1 that affect layer 2 attention patterns), we applied targeted attention removal to the layer 1 processing. The interaction between two objects is due to the attention computation in layer 1 where information from different patches converge. We investigated the contribution of the objected interactions by removing targeted interactions while keeping everything else intact. Specifically, for a given input game state, we constructed a manipulated game state where a set of objects, ie. the targeted objects, were removed. We then fed both the original and the manipulated game states into the transformer model and obtained the keys, queries, and values of layer 1 for each game state. Then for each patch, we compute layer 1 attention with keys generated from the original state, queries and values generated from the manipulated state. An patch's  attention to itself was still computed with the key, query and value from the original state. By doing this, the patches with the targeted objects could not receive information from other targeted objects, while everything else in the transformer worked normally. 

Mathematically, the manipulated attention $\overline{Attn}$ in layer \(1\) is computed as follows:



\begin{equation}
\overline{Attn}(x_m,\overline{\bf X})=a^{1}_{m,m}x_{m}{\bf W}^i_V+\sum_n{\overline{a}^{1}_{m,n\neq m}\overline{x}_{n}{\bf W}^i_V}
\end{equation}

\begin{equation}
    \overline{a}^{1}_{m,n}=\sigma(\frac{{x}_m {\bf W}_Q^1 {\bf W}_K^1 \overline{x}_n^{T}}{\sqrt{d_k}})\text{,}
\end{equation}

where \({\bf W}^1_V\) is the value weight matrix of the layer 1 attention block, \(a^1_{m,n}\) and \(\overline{a}^{1}_{m,n}\) are the original and manipulated attention score between the \(m\text{-}th\) patch and the \(n\text{-}th\) patch of layer 1, \(\overline{\bf X}\) is the embedding of the manipulated input, \(x_m\) and \(\overline{x}_m\) are the embeddings of the \(m\text{-}th\) patch of the original input and the manipulated input, \(\sigma\) represents the softmax function, respectively.

The result is then fed into the transformer and processed as normal. 

\subsection{Group-specific linear model}
To capture the relationship between patch attention and object attention, we leveraged the observation that the patch attention was linearly related to the sum of object attention and but discounted by the number of overlapping objects. We fit the following model by minimizing the mean squared error between the two sides:

\begin{equation}
   \alpha_{p}=\alpha_b+r^nq\sum_{i\in obj_p}\alpha_{o,i}\text{,}
\end{equation}

where \(\alpha_p\) is the patch attention, which is the raw attention score of a patch. \(\alpha_b\) is the spatial bias, defined as the raw attention score of a patch when all the none-maze channels are cleared from the input. \(obj_{p}\) is the set of objects in the current patch, and \(\alpha_{o,i}\) is the object attention of object \(i\), obtained by subtracting \(\alpha_b\) from the raw attention score when only object \(i\) is kept and all the other none-maze channels are cleared from the input. \(r\) and \(q\) are free parameters, which are the discount factor and the reference scaling factor, respectively. \(n\) is the total number of object overlaps in a patch. The number of overlaps of a tile is defined as the number of objects occupying the same tile subtracted by one, and the number of overlaps of a patch is the total overlaps of its four tiles.

\subsection{Shared object map}

Since each object within a patch contributes to the patch's attention independently, we may generate an attentional lookup map, where values on the map are the attention contributed by the object at the respective location. Therefore, we extracted these object attention maps by traversing all tiles and computing the attention of the patch to which the current tile belong, when only Pac-Man and the target object on the current tile are present. We then subtracted the computed attention from the baseline computed from an empty map. We averaged the object attention belonging to the same patch and assigned the averaged value to the tiles of this patch. 

To reveal the shared object map, we used this equation:

\begin{equation}
   M_o=c+u_oM\text{,}
   \label{eq shared map}
\end{equation}

where \(M_{o}\) is the attention map of object \(o\), \(c\), \(u_o\), and \(M\) are a constant, the scaling factor specific to each object, and the shared map, respectively. We solve this equation using an iterative algorithm, where \(u_o\) and \(M\) are alternately fixed while the other free parameters being optimized. The process is repeated until the result converges.

\subsection{Condensed model} 

We constructed the condensed normative model based on the computation abstracted from the transformer network (\ref{eq_condensed_moddel}). For the sake of simplicity, we used matrix forms to describe the model. We introduce a new matrix variable \(\mathbf{L}\) that indicates whether an objects is present or not on a tile. \(u\) is a stack of a vector of \(u_o\) in equation \ref{eq shared map} and a ones vector. \(\mathbf{M}\) consists of two column vectors: a latent map vector and a bias vector that corresponds to the \(c\) in equation \ref{eq shared map}.

To compare the condensed model with the original rollout score, we computed the Spearman correlation between the object-level rollout score and the attention predicted by the condensed model. 

To validate the model prediction in the monkeys, we showed the correlation between the predicted attention and the distance to the gaze positions of each tile. Since the reconstruction results in a lot of zero-attention tiles, we only show the top-n tiles that are non-zero in at least 10\% of the samples. 

\section{Acknowledgements}
This work was supported by the Strategic Priority Research Program of the Chinese Academy of Sciences (grant no. XDB1010301) and the National Science and Technology Innovation 2030 Major Program (grant no. 2021ZD0203701) to T.Y.. The funders had no role in study design, data collection and interpretation, or the decision to submit the work for publication.

\section{Author contributions}
T.Y. and Z.L. conceptualized the study. Z.L. and Y.L performed the experiments and collected behavioral data. Z.L. performed the computational modelling and analyses. T.Y. and Z.L. wrote the manuscript.

\section{Competing interests}
The authors declare no competing interests.

\newpage
\bibliographystyle{plainnat}
\bibliography{neurips_2024}


\clearpage
\appendix
\label{app:appendix}
\nolinenumbers 

\vspace*{10cm}
{\bfseries \Huge Supplementary Information}

\clearpage
\begin{table}[hbtp]
  \caption{Channel Descriptions of the Model Input}
  \label{channels}
  \centering
  \begin{tabular}{c c p{8cm} }
    \toprule
      Channel Number & Channel Name & Description\\
    \midrule
        1 & maze & 1 if the tile is a wall tile, 0 if the tile is a walkable tile.\\
    \midrule
        2 & bean & 1 if a bean is on the tile, 0 otherwise.\\
    \midrule
        3 & energizer & 1 if an energizer is on the tile. Eating an energizer leads to 4 drops of juice reward and turns the ghosts into the scared mode.\\
    \midrule
        4 & apple & 1 if an apple is on the tile. Taking an apple leads to 12 drops of juice reward.\\
    \midrule
        5 & cherry & 1 if a cherry is on the tile. Eating a cherry leads to 3 drops of juice reward.  \\
    \midrule
        6 & melon & 1 if a melon is on the tile. Eating a melon leads to 17 drops of juice reward.  \\
    \midrule
        7 & orange & 1 if an orange is on the tile. Eating an orange leads to 8 drops of juice reward.  \\
    \midrule
        8 & strawberry & 1 if a strawberry is on the tile. Eating a strawberry leads to 5 drops of juice reward.  \\
    \midrule
        9 & ghost 1 (normal) & 1 if ghost 1 in normal mode is on the tile. Colliding into ghosts in normal mode leads to a time-out penalty. After the time-out, Pac-Man respawns at the start tile and continues the previous game.\\
    \midrule
        10 & ghost 2 (beaten) & 1 if ghost 1 in beaten mode is on the tile. A beaten ghost moves back to the ghost house. Colliding into a beaten ghost does not trigger any game events.\\
    \midrule
        11 & ghost 1 (scared) & 1 if ghost 1 in scared mode is on the tile. Eating a scared ghost turns the ghost into the beaten mode and produces 8 drops of juice reward.\\
    \midrule
        12 & ghost 1 (flashing) & 1 if ghost 1 in flashing mode is on the tile. A flashing ghost is a scared ghost that will turn back into the normal model very soon. \\
    \midrule
        13 & ghost 2 (normal) & 1 if ghost 2 in normal mode is on the tile. \\
    \midrule  
        14 & ghost 2 (beaten) & 1 if ghost 2 in beaten mode is on the tile. \\
    \midrule
        15 & ghost 2 (scared) & 1 if ghost 2 in scared mode is on the tile. \\
    \midrule
        16 & ghost 2 (flashing) & 1 if ghost 2 in flashing mode is on the tile. \\
    \midrule
        17 & Pac-Man & 1 if Pac-Man is on the tile, 0 otherwise  \\
    \bottomrule
  \end{tabular}
\end{table}

\setcounter{figure}{0}
\renewcommand{\thefigure}{S\arabic{figure}}

\clearpage
\begin{figure}[t]
    \hspace{0.\textwidth}
        \includegraphics[width=1.05\textwidth]{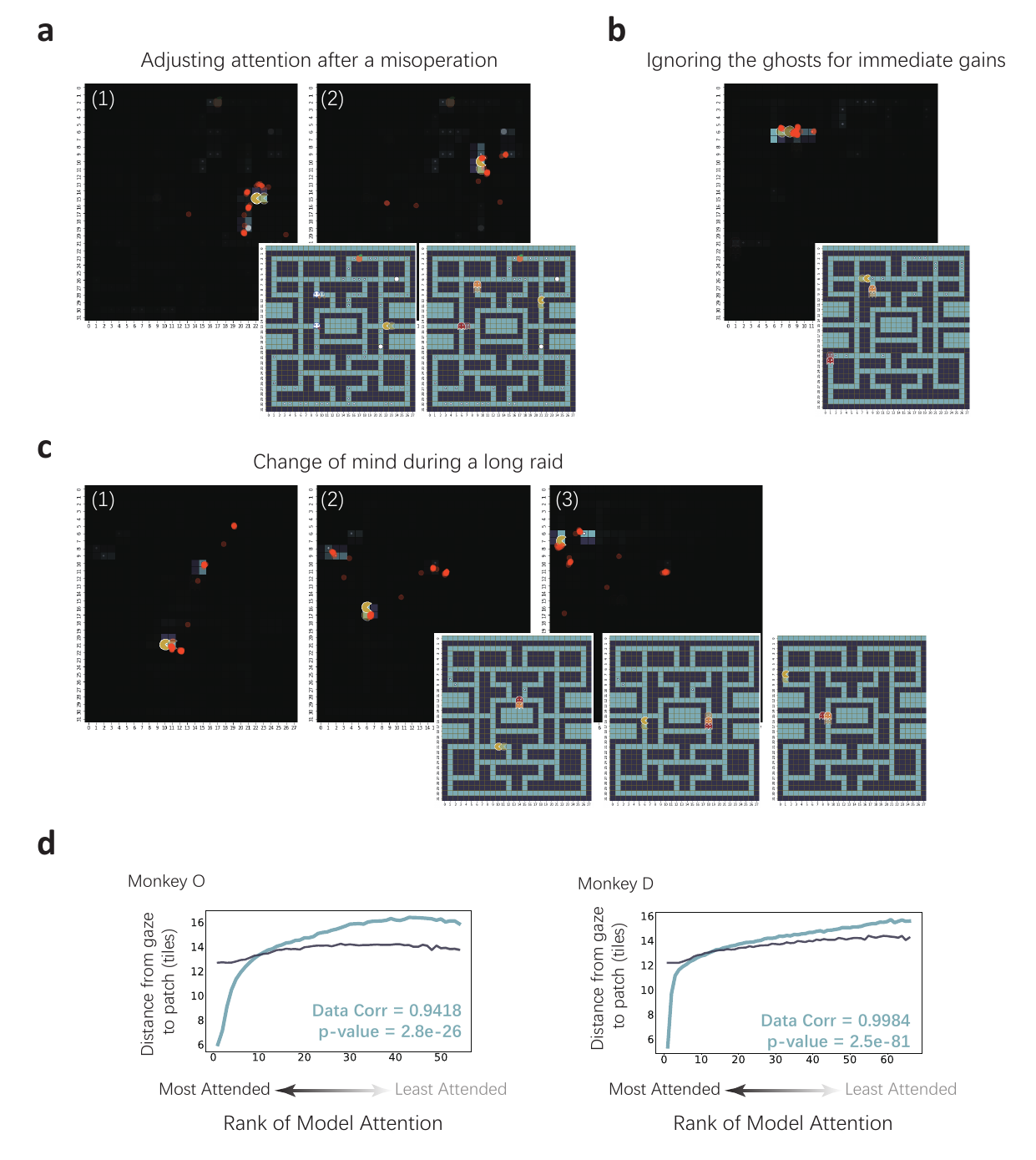}
    \caption{\textbf{More examples and statistics supporting the correlation between rollout score and gaze pattern. a.} The monkey was initially attending to the energizer on the bottom branch. However, it swiftly redirected its attention toward the beans on the upper branch following a potentially erroneous action. \textbf{b.} The monkey ignored an approaching ghost in proximity for immediate bean rewards. \textbf{c.} The monkey was attending to a bean on the central axis of the maze initially and gradually changed its focus to the beans on the upper left corner. The rollout score captured the shift of attention even before the Pac-Man moved toward a different direction than originally planned. \textbf{d.} The patch's attention score rank order is correlated with the distance of the monkeys' gaze location from the patch where at least one object is present (top: monkey O; bottom: monkey D). The light blue curve is the session average, the error bars (standard error of the mean) are too small to be visible. The dark blue is the random control, where the rank order of non-empty patches in each sample is shuffled.}
    
\label{fig:more example}
\end{figure}

\clearpage
\begin{figure}[t]
    \hspace{0.\textwidth}
        \includegraphics[width=1.05\textwidth]{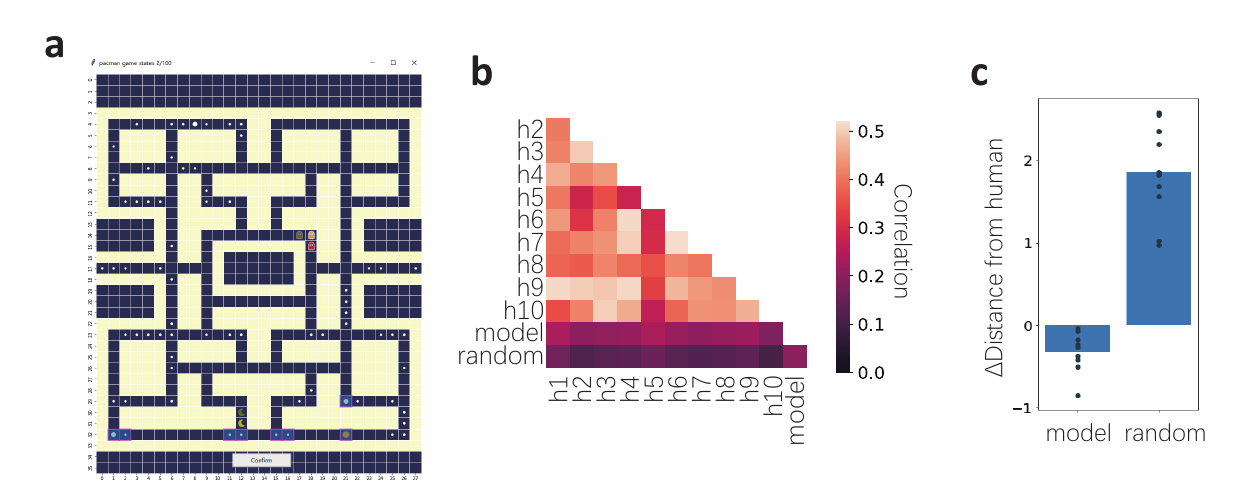}
    \caption{\textbf{Human experiments. a.} An example experiment screenshot. Ten healthy human subjects (two male, eight female; age, 24-32) participated in the experiment. During the experiment, they sat in front of a monitor and watched 100 game states that were randomly chosen from monkey O's dataset. The participants were asked to choose 1-10 tiles that were most relevant to the impending decision at the next junction by clicking the mouse.  There was no set time limit for the task, and the participants finished the experiment using 30-55 minutes. \textbf{b.} Jaccard similarities (\(J(A,B) = |A\cap B|/|A\cup B|\)) between the human subjects' judgements, the attention rollout scores, and a random control, which assigns attention randomly to the objects in the maze. The number of selected tiles of both the model and the random control were matched to the human subject's choice that was being compared against. The humans made similar judgements (bright color), but differed from the network's attention rollout scores. \textbf{c.} Compared to the human baseline, the prediction based on the attention rollout score had a smaller error. The random control had much greater errors. A positive difference indicates greater error and worse prediction performance compared to the human. Blue bars are the average differences across all subjects. For each agent, the prediction error was quantified by first assigning the nearest selected tile to each gaze data point and then averaging the distance between the assigned tiles and the gaze position. To compare the prediction accuracy between the models and the participants, we subtract the models' prediction error against the participants'. A negative difference indicates better model performance. Note that this analysis is sensitive to the number of selected tiles, so we matched the number of tiles for each comparison. We also observed that human participants tended not to choose the Pac-Man tile. This was most likely because the instruction given to the human subjects encouraged them not to include Pac-Man in their answers.  Both the monkeys and the transformer model assigned a large amount of gazes or attention to Pac-Man. Therefore, we included the Pac-Man tile in the human responses when evaluating human-model differences.}
\label{fig:hum exp}
\end{figure}

\clearpage
\begin{figure}[t]
    \hspace{0.\textwidth}
        \includegraphics[width=1.05\textwidth]{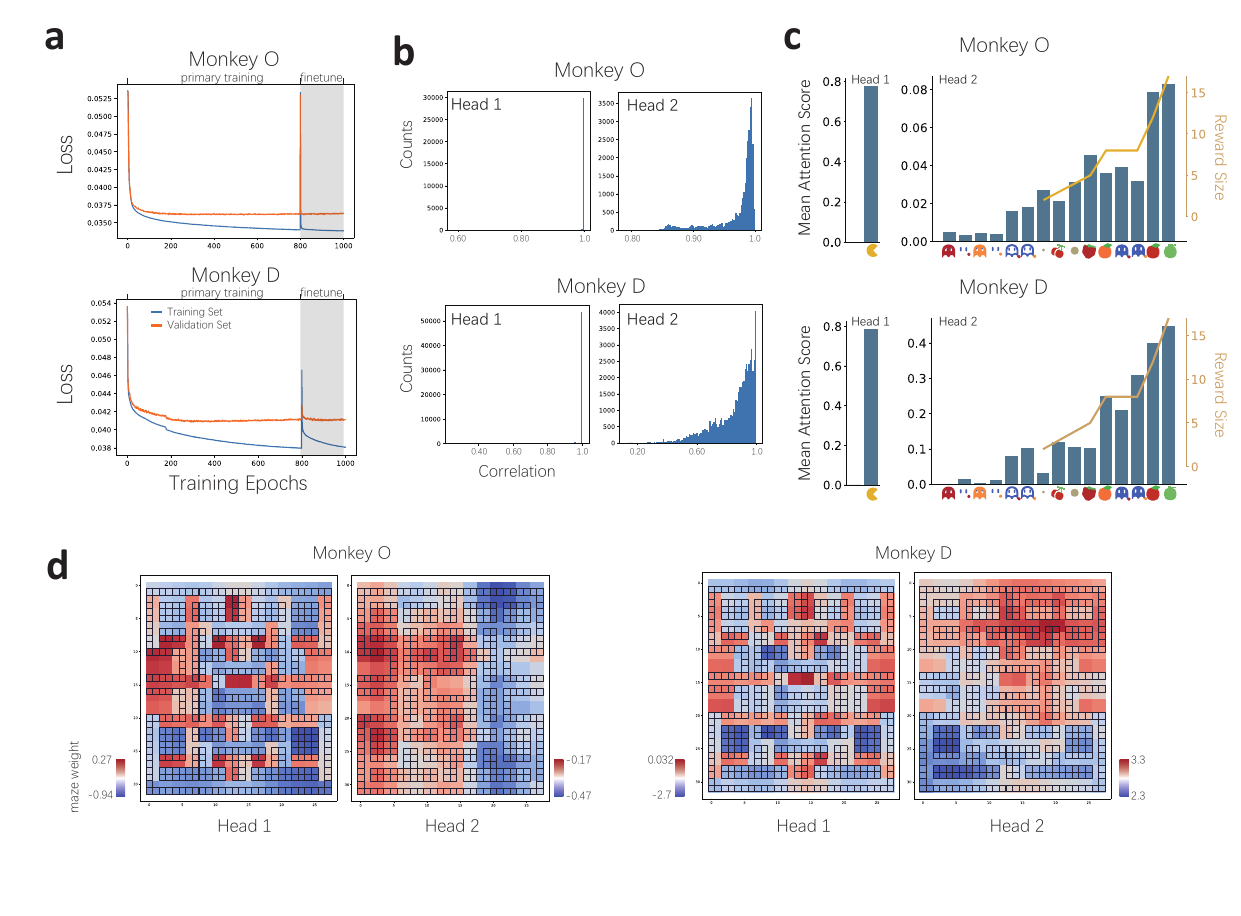}
    \caption{\textbf{Finetuning to remove the layer norm computation. a.} The loss curve during training. After removing the first layer norm computation and starting the finetune protocol, both the training loss and validation loss first increased drastically and then decreased to the stable level before finetune within 200 epochs. \textbf{b.} The attention pattern of the finetuned model is highly correlated with the orriginal model, making the finetuned model a good proxy for studying the computation underlying the layer 1 attention. \textbf{c.} Mean attention scores of the original model computed across test samples. mean attention scores of head 2 (left y-axis) correlate with the objects' value (right y-axis).  \textbf{d.}  Maze-component of the finetuned model.}
\label{fig:finetuning}
\end{figure}
\clearpage

\clearpage
\begin{figure}[t]
    \hspace{0.\textwidth}
        \includegraphics[width=1.05\textwidth]{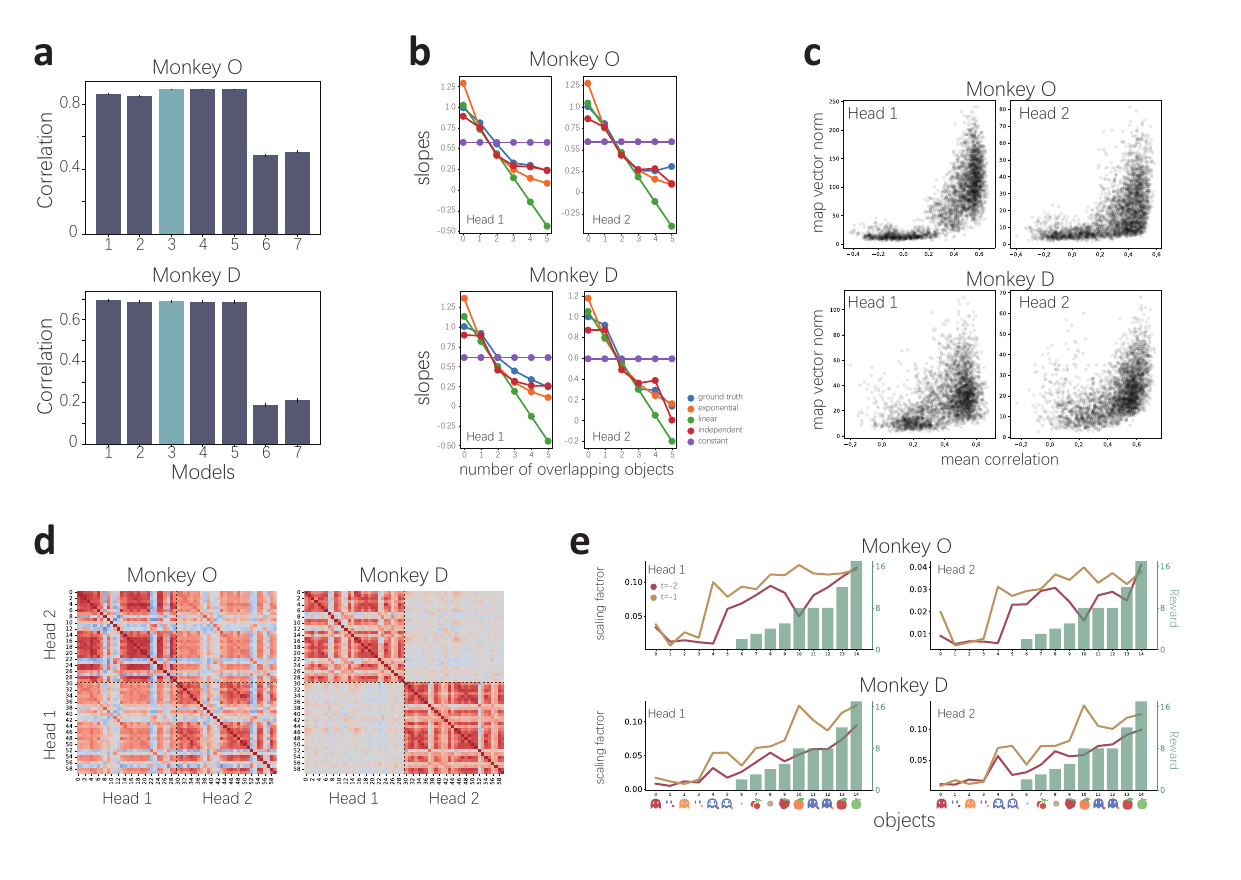}
    \caption{\textbf{The computation mechanism underlying layer 2 attention. a.} The performance (depicted as the Spearman correlation between the patch attention and the prediction) of 7 candidate models of patch attention. The models are listed by the number of free parameters: 1) constant slope model: \(h(n)=r_0\sum{Attn_{obj}}+b\); 2) maximum object attention model: \(h(n)=r_0(Attn_{obj})_{max}+b\), where \((Attn_{obj})_{max}\) outputs the maximum object attention in a patch; 3) exponential slope model: \(h(n)=r_0r^n\sum{Attn_{obj}}+b\); 4) linear slope model: \(h(n)=(r_0-nr)\sum{Attn_{obj}}+b\); 5) independent slope model:  \(h(n)=r_n\sum{Attn_{obj}}+b\), where \(r_n\) is a set of slope parameters that depend on the number of overlaps; 6) object-dependent model:  \(h(n)=\sum{r_{obj}Attn_{obj}}+b\); and 7) soft-prioritizing model: \(softmax(R_{obj})*Attn_{obj}+b\), where \(R_{obj}\) is an learnable object-dependent rating score that determines the priority rankings between objects. \textbf{b.} Fitting the 4 models that account for the slope of the patch attention model generates different results. The exponential slope model and the independent slope model provide closer fittings to the ground truth. \textbf{c.} The vector norm of object attention maps is plotted against the mean correlation with all object attention maps given the same Pac-Man location. Each dot represents an object given a Pac-Man location. The point clouds can be roughly divided into two clusters, which can be characterized as a low-correlation and low-contribution cluster, and a high-correlation and high-contribution cluster, respectively. \textbf{d.} The correlation of object attention maps across the two heads. The two attentional heads of monkey O's model share similar patterns, while monkey D's are less correlated, suggesting that monkey D's goal-directed attention might consist of more complicated structures. \textbf{e.} The map scaling factor of each object is correlated with the object reward. }
\label{fig:layer 2 supplementary}
\end{figure}

\clearpage
\begin{figure}[t]
    \hspace{-0.2\textwidth}
        \includegraphics[width=1.4\textwidth]{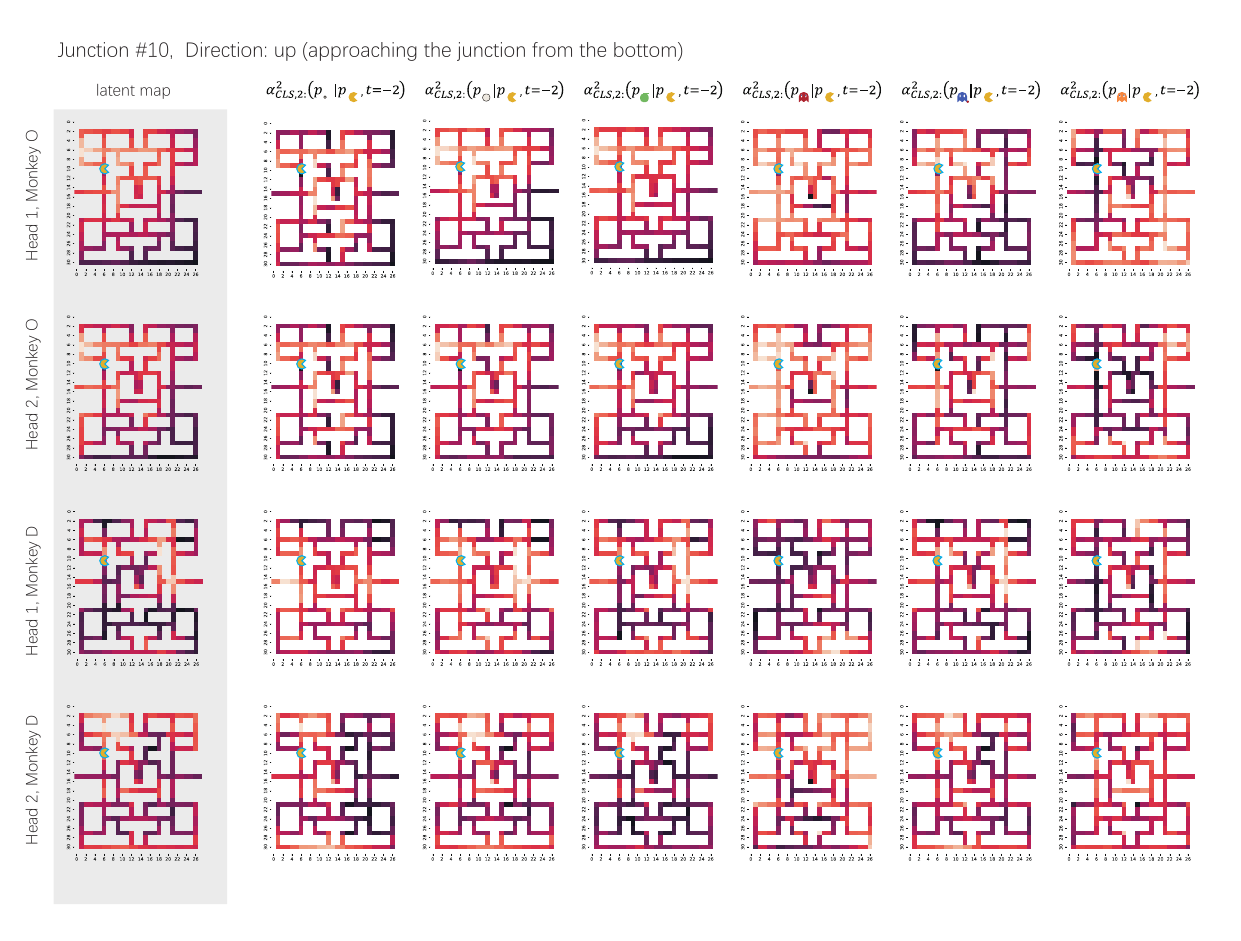}
    \caption{\textbf{Example object attention maps.} The left-most column shows the fitted shared object attention map. The object-specific attention map of 6 objects (column) are shown on the right for each head of the layer 2 in the tow monkeys (row).}
\label{fig:example weight map}
\end{figure}

\clearpage
\begin{figure}[t]
    \hspace{-0.2\textwidth}
        \includegraphics[width=1.4\textwidth]{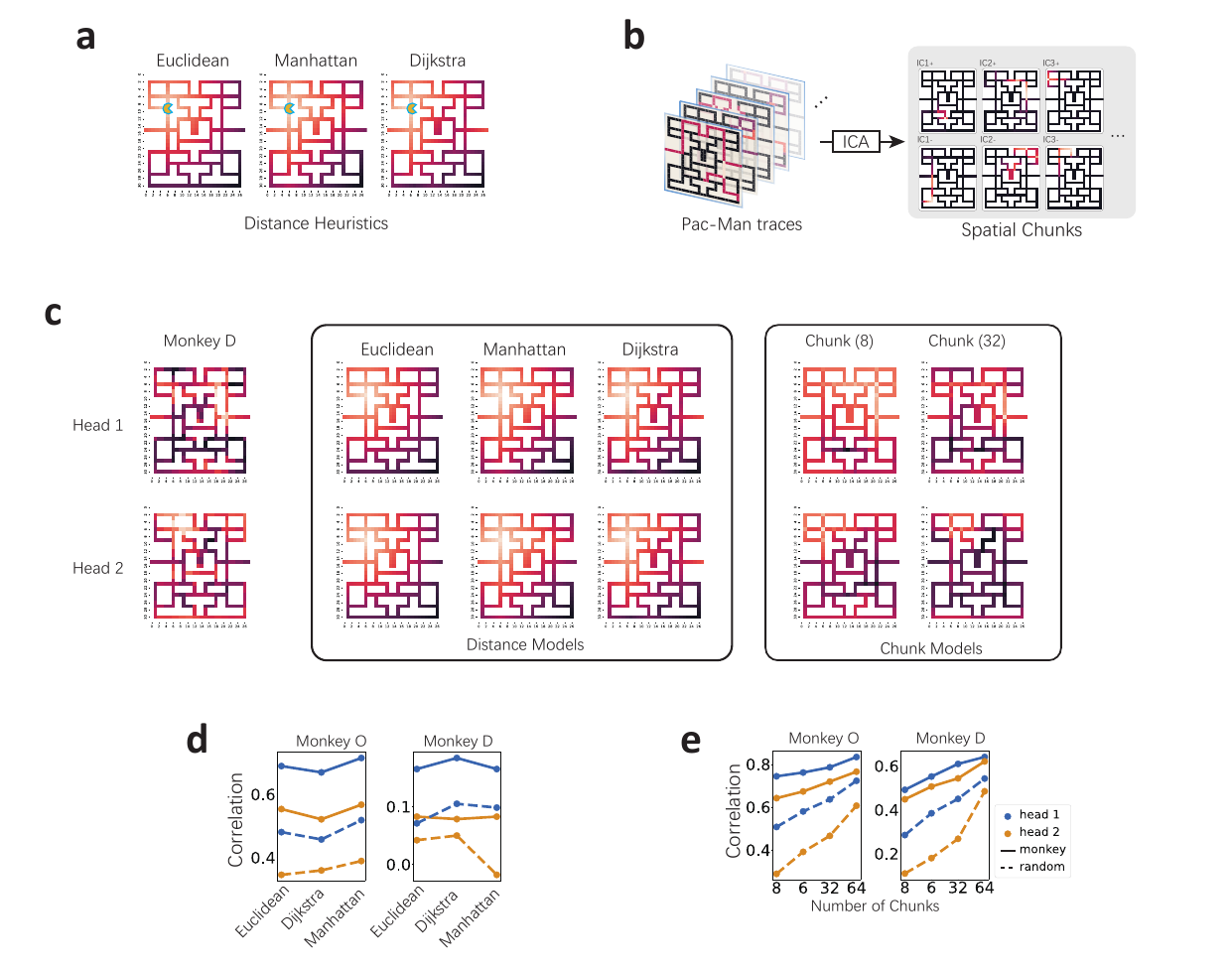}
    \caption{\textbf{Heuristic models of latent map. a.} shows three examples of distance heuristics: the Euclidean distance, the Manhattan distance, and the the Dijkstra distance, respectively. \textbf{b.} is schema indicating how spatial chunk heuristics are generated. We extract actual Pac-Man traces generated during gameplay and crop them into sequences of 100 tiles. Different length of sequences were tested and did not affect the results. We then apply an independent component analysis on the traces and the independent components are the chunk heuristics. \textbf{c.} shows an examples of reconstructed latent map from each model. The first column is the latent map of monkey D. Three distance models compute linear regressions between the latent map and the Euclidean distance, Manhattan distance, and Dijkstra distance, respectively. The chunk models compute multivariate linear regression between the latent map and the independent components. The number in the brackets indicates the number of components, which is a hyperparameter of independent component analysis. \textbf{d.} The correlation between the distance model reconstruction and the latent map. The solid lines represent actual data and the dashed lines represent a random control where the distance heuristics are shuffled across junctions. \textbf{e.} The correlation between the chunk model reconstruction and the latent map. More chunks lead to higher reconstruction performance.
    }
\label{fig:latent map models}
\end{figure}
\clearpage

\end{document}